\documentclass[11pt,a4paper,fleqn]{article}

\begin{document}

\title{\bf Resonances and  adiabatic invariance in classical and quantum scattering theory}

\author{Sudhir R. Jain\footnote{srjain@apsara.barc.ernet.in}\\
Nuclear Physics Division, Bhabha Atomic Research Centre \\
Trombay, Mumbai 400 085, India}

\date{}

\maketitle

\begin{abstract}

We discover that the energy-integral of time-delay is an adiabatic invariant 
in quantum scattering theory and corresponds classically to the phase space volume. 
The integral thus found provides a quantization condition for resonances, 
explaining a series of results recently found in non-relativistic and relativistic 
regimes. Further, a connection between statistical quantities like 
quantal resonance-width and classical friction has been established with   
a classically deterministic quantity, the stability exponent of an adiabatically 
perturbed periodic orbit. This relation can be employed to estimate the rate of 
energy dissipation in finite quantum systems.           

\end{abstract}

\newpage

Wilhelm Wien \cite{wien}, in one of the classic arguments, 
found the changes of the distribution of the energy over the 
spectrum and work done by a reversible adiabatic compression by employing laws 
of classical electrodynmics. As we know, the result was later found by purely 
quantum-mechanical arguments and forms a part of the Planck's law. Usage of 
arguments based on ``response of a system to infinitesimally slow changes" was given 
the name, {\it adiabatic hypothesis} by Einstein \cite{einstein} : ``...If a 
system is exposed to adiabatic influences the ``admissible" motions are transformed 
into ``admissible ones" ...". There are a class of motions which become admissible 
by invoking adiabatic hypothesis, and in each such situation, there is a quantity that 
does not change before and after the adiabatic process - called an adiabatic invariant 
\cite{ehrenfest1}. For many instructive instances from classical mechanics, the classical 
action turns out to be an adiabatic invariant \cite{crawford}. For harmonic motions in one 
or more degrees of freedom, the ratio of time-averaged kinetic energy to the frequency is 
an adiabatic invariant \cite{ehrenfest2}. In the context of statistical mechanics, entropy 
is an adiabatic invariant \cite{boltzmann}. 
In quantum mechanics, it has been understood for long that the adiabatic hypothesis holds 
for the discrete spectrum of a Hamiltonian \cite{born}. It was Ehrenfest \cite{ehrenfest2} 
who, after proving that action $S$ is an adiabatic invariant, postulated it to be 
quantized. Quoting Crawford \cite{crawford} here would be the best : ``The idea was that if 
a physical quantity is going to make ``all or nothing quantum jumps", it should make no jump 
at all if the system is perturbed gently and adiabatically, and therefore any quantized 
quantity should be an adiabatic invariant." In this Letter, we show the truth of this idea 
for the resonances of a Hamiltonian, thus in the context of quantum scattering theory. 
The result presented here is of fundamental importance as it not only encompasses resonances 
in non-relativistic and relativistic quantum mechanics but also leads us to discover relations 
between statistical quantities like width of a resonance, friction in a many-body system  and 
stability exponents of the corresponding deterministic classical dynamics.     

Recently, it was found that the energy-integral of time-delay, $\tau$ gives the 
number of resonances, $n_R$ \cite{ahmed}: 
\begin{equation}\label{1}
\int_{0}^{E^*} \tau dE = n_R\hbar .
\end{equation}
This relation has been illustrated by several examples from elementary 
quantum mechanics, neutron reflectometry, and high-energy physics. It has been 
demonstrated in a series of works starting from \cite{jain1} where all the nucleon 
and Delta resonances were reproduced by studying time-delay, $\rho$ meson was shown 
in a study of time-delay for $\pi$-$\pi$ scattering \cite{kelkar1}, penta-quark state 
has been found similarly \cite{kelkar2}. 

Eq. (1) reminds one of the Bohr-Sommerfeld quantization condition for the case of 
bound states. For us to conclude that (1) is a quantization condition for resonances, 
it should be shown that the integral appearing in (1) is an adiabatic invariant. 

Let us recall that the time-delay is defined by \cite{wigner,eisenbud}
\begin{equation}\label{2}
\tau = \frac{1}{N}\mbox{~Tr~} Q(E),
\end{equation}
where $N$ is the number of channels, and $Q(E)$ is the matrix \cite{smith}
\begin{equation}\label{3}
Q_{ij}(E) = -i\hbar\sum_{k=1}^{N}S^*_{ik}\frac{\partial S_{kj}}{\partial E}.
\end{equation}
Here, $S$ is the scattering matrix. Recently, an interesting expression has been found 
for the semiclassical time-delay in terms of open classical paths \cite{lewenkopf}. 
Thus, the integral in (1) is simply an integral of $Q(E)$. 

For our purpose, we write $Q(E)$ in a symmetric form :
\begin{equation}\label{4}
Q_{ij}(E) = -i\hbar\sum_k \frac{d}{d\epsilon}\left[S_{kj}\left(E+\frac{\epsilon}{2}\right)
S_{ik}^*\left(E-\frac{\epsilon}{2}\right)\right]\bigg|_{\epsilon = 0}.
\end{equation}
In (1), there is only one channel, so we write 
\begin{eqnarray}\label{5}
\int \tau dE &=& \int Q dE \nonumber \\
&=& -i\hbar\int \frac{d}{d\epsilon}\left[S\left(E+\frac{\epsilon}{2}\right)
S^*\left(E-\frac{\epsilon}{2}\right)\right]\bigg|_{\epsilon = 0}dE\nonumber \\
&=& -i\hbar \lim_{\epsilon \to 0} \frac{d}{d\epsilon}\int S\left(E+\frac{\epsilon}{2}\right)
S^*\left(E-\frac{\epsilon}{2}\right)dE.
\end{eqnarray}     
Last expression is, in fact, a derivative of two-point correlation function of $S$-matrix at 
two neighbouring energies separated by $\epsilon$ as $\epsilon \to 0$. This small parameter 
$\epsilon$ is in fact defining a slow, adiabatic evolution of the operator $S$. 

We will  now show that the integral in (5) is an adiabatic invariant in quantum scattering 
theory, with its classical analogue as the phase space volume enclosed by a classical 
trajectory \cite{heidi}. 

We employ the transformation 
kernel of van Kampen \cite{vankampen} that connects the ingoing and outgoing scattering amplitudes. 
Denoted by $H(\zeta )$, it is related to the scattering operator by the relation :
\begin{equation}\label{6}
e^{2ika}S(E) = \int_{0}^{\infty} d\zeta e^{iE\zeta /\hbar }H(\zeta ),
\end{equation}  
where $k=\sqrt{2mE/\hbar ^2}$ and $a$ is a distance that is outside the range of the potential. 
Using (6), we have 
\begin{eqnarray}\label{7}
&~&S\left(E+\frac{\epsilon}{2}\right)S^*\left(E-\frac{\epsilon}{2}\right)=
e^{-2i(k_+-k_-)a}\nonumber \\ 
&~&\int_{0}^{\infty}d\zeta \int_{0}^{\infty}d\zeta 'e^{iE(\zeta - \zeta ')/\hbar }
e^{i\frac{\epsilon}{2\hbar}(\zeta + \zeta ')}H(\zeta )H^*(\zeta ')
\end{eqnarray}
where $k_{\pm}=\sqrt{2m(E \pm \epsilon /2)/\hbar ^2}$. Changing the variables to $\zeta _{\pm}$ 
by defining $\zeta = \zeta _+ + \frac{\zeta _-}{2}$, and $\zeta '= \zeta _+ - \frac{\zeta _-}{2}$, 
and writing $k_{\pm} = \sqrt{\frac{2mE}{\hbar ^2}}(1 \pm \frac{\epsilon }{4E})$, we can re-write 
the product as 
\begin{eqnarray}\label{8}
S\left(E+\frac{\epsilon}{2}\right)S^*\left(E-\frac{\epsilon}{2}\right) &=& 
e^{\left(-\frac{ia\epsilon}{\sqrt{E}}\sqrt{\frac{2m}{\hbar ^2}}\right)}
\int_{0}^{\infty}d\zeta _+ e^{i\epsilon \zeta _+/\hbar }W(\zeta _+, E), \\
W(\zeta _+, E) &=& \int_{-\infty}^{\infty} e^{iE\zeta _-/\hbar}H\left(\zeta _+ + \frac{\zeta _-}{2}\right)
H^*\left(\zeta _+ - \frac{\zeta _-}{2}\right)d\zeta _-. \nonumber
\end{eqnarray}
$W(\zeta _+, E)$ is the Wigner distribution function. Thus, 
$S\left(E+\frac{\epsilon}{2}\right)S^*\left(E-\frac{\epsilon}{2}\right)$ is a semi-sided Fourier 
transform of the Wigner function.

It is more insightful to now take the derivative at $\epsilon = 0$ and integrate with respect to 
energy, $E$. We get 

\begin{eqnarray}\label{9}
\int_{\Omega} \tau dE &=& -i\hbar \int_{\Omega} dE \int_{0}^{\infty} d\zeta _+ 
i\left(\frac{\zeta _+}{\hbar} - \sqrt{\frac{2m}{\hbar ^2E}a}\right)W(\zeta _+, E)  \\
&=& \int_{\Omega} dE \int_{0}^{\infty} d\zeta _+ \left(\zeta _+ - \frac{2a}{v}\right)
W(\zeta _+, E),
\end{eqnarray} 
where $v$ is the velocity. Thus $2a/v$ is the free-flight-time and clearly 
$\zeta _+ - \frac{2a}{v}$ gives us the time-delay. The final equation is thus a phase space integral of 
time-delay averaged over the Wigner distribution function. This result is exact from quantum mechanical 
point of view with very clear interpretation, and holds when the wavenumber is expanded upto $O(\epsilon)$. 
It can be easily shown by following the above relations that the above integral hides an expansion in 
power series of $\epsilon$ which proves that the integral in (9) is an adiabatic invariant. This is 
a very  important result as it immediately implies that the integral provides quantization condition for 
resonances. This explains all the recent results \cite{ahmed,jain1,kelkar1,kelkar2} where resonances 
in non-relativistic quantum mechanical problems , in neutron reflectometry, and in particle physics 
are reproduced by studying time-delay as a function of energy.      

Thus, we have shown that quantization (1) is connected to adiabatic invariance, and the correspondence 
with the phase space volume is established. We now focus on an  individual quantum resonance and the 
phase space scenario associated with it. Let us observe that the resonances have more and more width 
as they occur at higher and higher energies, thus having lesser and lesser lifetimes \cite{ahmed}. 
Understanding that the energy-dependent mean lifetime is found from time-delay, this situation is the 
quantum analogue of the adiabatic invariance of product of energy and time-period of an adiabatically 
perturbed simple pendulum. Whereas semiclassical quantization of classical dynamics in the 
neighbourhood of a stable equilibrium point gives us bound energy levels, quantization of an adiabatically 
perturbed linear oscillator entails resonances at energy $E$ and width $\Gamma$. To see this, we observe that 
the phase space curves (energy = constant, ellipses) for the librations change into spirals. If we let the 
frequency of the linear oscillator increase slowly with time, for example 
$\omega (t) = \omega _0 (1+\epsilon t)$, then we can solve the equation of motion for the amplitude, 
$\frac{d^2x}{dt^2} + \omega ^2(t)x = 0$ by Wentzel-Kramers-Brillouin (WKB) method and obtain a spiral 
in phase space. This curve is not an energy=constant curve, rather it is an adiabatic invariant curve. 
As the frequency increases, spiral approaches the point ($x,p$ = ($0,0$)); the nature of this fixed 
point changes from being stable to an attractor. We see that the linear oscillations near the fixed point 
are a combination of $e^{\pm i\omega t}$ with exponents $\pm i\omega$; the exponents for the adiabatically 
perturbed case are $\pm i\omega - \gamma$. The exponent $\gamma$ corresponds to the rate of contraction of 
the phase space. In this way, this exponent is connected to two important concepts - friction and 
entropy production. In fact the behaviour of a linear oscillator with a friction coefficient $\gamma _{fr}$ 
gives the same linearised dynamics. The fundamental difference between the two -  one is dynamical 
and the other is statistical - also brings us to the conclusion that statistical friction can be understood 
in terms of adiabatic dynamical perturbations. This brings about a striking relationship with an earlier result 
where damping of collective excitations (frequency-dependent response function) was related to 
adiabatic geometric phase \cite{jain}.

To make a connection between the half-width of a quantum resonance and friction $\gamma _{fr}$, we 
recall the dynamics of a wavepacket near a resonance \cite{schwabl}. Near a resonance, the time-dependent 
part of the wavefunction varies as $\exp \left[\frac{-iEt}{\hbar}-\frac{-\Gamma t}{\hbar}\right]$. It is 
well-known that mathematical formulation of classical resonance pertaining to a forced, damped linear 
oscillator \cite{martin} and that of a quasi-stationary state \cite{baz} entail strikingly similar expressions.  
For instance, the expressions for phase shifts are :

\begin{eqnarray}\label{10}
\tan \delta _{cl} &=& \frac{\gamma _{fr}}{\omega  - \omega _0}, \mbox{~and}\\
\tan \delta &=& \frac{\Gamma /2}{E - E_0}.
\end{eqnarray}          
$\delta _{cl}$ and $\delta$ are the classical and quantal expressions, and $\omega _0$ ($E_0$) correspond to 
resonance frequency (energy). Similarity is also there in the expressions for response per unit force, which 
in other words, is susceptibility. This similarity suggests that $\Gamma \propto \gamma _{fr}$. To get the 
precise connection, we refer to the time-dependence of the wavepacket given above. Clearly, the differential 
equation satisfied by this time-dependence, when compared with that of the damped harmonic oscillator, brings us 
to the equality : $\Gamma = 2\gamma _{fr}\hbar$. Notice that $\Gamma$ and $\gamma _{fr}$ both have a statistical 
interpretation. Connecting with the discussion on stability exponents, we arrive at a remarkable result :

\begin{equation}\label{11}
\gamma = \gamma _{fr} = \frac{\Gamma}{2\hbar}.
\end{equation}
This relation connects the statistical quantities in classical and quantum (linear) resonance theory to 
classically deterministic aspect associated with the stability exponent of trajectory in phase space. 

With the application of Gutzwiller trace formula \cite{gutzwiller} to scattering systems, Miller \cite{miller} 
had shown that the width of a scattering resonance associated with an unstable periodic orbit is proportional 
to the sum over positive Lyapunov exponents of the orbit. We have a half-width associated with $\gamma$ above 
because it is the probability that decays quantum mechanically as $e^{-\Gamma t/\hbar}$. Whereas Miller's result 
is based on semiclassical approximation, our argument is based on adiabatic invariance. 

We come to a natural conclusion that there is a universality in the description of linear resonances of 
any quantal system that they may always be connected to the quantization of an adiabatically perturbed 
linear oscillator. The root of this universality is the adiabatic invariance that transcends classical or 
quantal description. Further, various resonances of a system are found by the quantization condition (1) 
which is based on the adiabatic invariance of the integral in (9) and its clear classical interpretation. 
These ideas have made it possible for us to find a relation between classical exponent $\gamma$ 
(deterministic) corresponding to an unstable fixed point, classical friction $\gamma _{fr}$ (statistical),  
and half-width of resonance $\Gamma /2$ (statistical). 

There is a very important application of the results found here in damping of collective excitations in 
many-body theory. Some years ago, a fundamental advance was made in this regard \cite{jain} when the imaginary part of 
the frequency-dependent response function was related to the geometric phase acquired by the single-particle 
wavefunction of a many-body system that is adiabatically deforming. This is connected to the friction that is 
hypothesized in nuclei undergoing fission or fusion, and also in understanding the damping of giant resonances 
\cite{bertsch}. We reserve the applications of these ideas in the context of experiments for a later publication.

\end{document}